\documentclass[final]{emulateapj}
\usepackage{apjfonts}
\usepackage{amssymb,latexsym,amsmath,graphics}
\slugcomment{} \shorttitle{QPOs AS GLOBAL MODES IN THE BOUNDARY
LAYERS OF ACCRETION DISKS}
\shortauthors{ERKUT, PSALTIS, AND ALPAR}

\begin{document}

\title{Quasi-periodic oscillations as global hydrodynamic modes in the boundary layers of viscous accretion disks}

\author{M. Hakan Erkut,\altaffilmark{1,2} Dimitrios Psaltis,\altaffilmark{2,3} and M. Ali Alpar\altaffilmark{3}}

\affil{\altaffilmark{1}Department of Mathematics and Computer
Science, \.Istanbul K\"{u}lt\"{u}r University, Atak\"{o}y Campus,
Bak\i rk\"{o}y 34156, \.Istanbul, Turkey}

\affil{\altaffilmark{2}Physics
Department, University of Arizona, 1118 E. 4th St., Tucson, AZ
85721}

\affil{\altaffilmark{3}Faculty of Engineering and Natural
Sciences, Sabanc\i\ University, 34956, Orhanl\i, Tuzla, \.Istanbul,
Turkey}

\altaffiltext{1}{m.erkut@iku.edu.tr}
\altaffiltext{2}{dpsaltis@physics.arizona.edu}
\altaffiltext{3}{alpar@sabanciuniv.edu}

\begin{abstract}
The observational characteristics of quasi-periodic oscillations
(QPOs) from accreting neutron stars strongly indicate the
oscillatory modes in the innermost regions of accretion disks as a
likely source of the QPOs. The inner regions of accretion disks
around neutron stars can harbor very high frequency modes related to
the radial epicyclic frequency $\kappa $. The degeneracy of $\kappa
$ with the orbital frequency $\Omega $ is removed in a non-Keplerian
boundary or transition zone near the magnetopause between the disk
and the compact object. We show, by analyzing the global
hydrodynamic modes of long wavelength in the boundary layers of
viscous accretion disks, that the fastest growing mode frequencies
are associated with frequency bands around $\kappa $ and $\kappa \pm
\Omega $. The maximum growth rates are achieved near the radius
where the orbital frequency $\Omega $ is maximum. The global
hydrodynamic parameters such as the surface density profile and the
radial drift velocity determine which modes of free oscillations
will grow at a given particular radius in the boundary layer. In
accordance with the peak separation between kHz QPOs observed in
neutron-star sources, the difference frequency between two
consecutive bands of the fastest growing modes is always related to
the spin frequency of the neutron star. This is a natural outcome of
the boundary condition imposed by the rotating magnetosphere on the
boundary region of the inner disk.
\end{abstract}


\section{Introduction\label{intr}}

Wave modes in the boundary region in the inner accretion disk are a
likely source of the distinct narrow frequency bands of
quasi-periodic oscillations (QPOs) (van der Klis 2000) from neutron
star sources in low mass X-ray binaries (LMXBs). The dependence of
QPO frequencies on accretion rate suggests that the observed QPOs
are connected with accretion disk modes. Psaltis, Belloni \& van der
Klis (1999) showed the existence of correlations between the
frequencies of different QPO bands extending over a wide span of
frequencies. The same correlation encompasses black hole as well as
white dwarf and neutron star sources. This strongly implies that the
frequency bands are determined by the oscillation modes of the
accretion disk. The nature of the compact object, whether it is a
black hole, white dwarf or neutron star, probably plays a role in
exciting the same disk modes through possibly different mechanisms
in different types of compact sources. The QPO bands modifying the
X-ray luminosity are likely to belong to the boundary region between
the disk and the compact object. For magnetic neutron stars, this
boundary region is shaped by the interaction of the disk with the
magnetosphere. The region where the rotation deviates from Keplerian
flow is not necessarily very narrow; this boundary zone may have a
size as large as a few tenths of the inner disk radius. We shall
therefore use the terms \textquotedblleft boundary\textquotedblright
\thinspace and \textquotedblleft transition\textquotedblright
\thinspace region interchangeably.

Initial explorations of disk modes underlying QPO frequency bands
from LMXBs were provided by Alpar et al. (1992) and Alpar \& Y\i
lmaz (1997). The characteristics and possible excitation mechanisms
of thin-disk oscillations were reviewed and discussed by Kato
(2001). In many current models of QPOs, especially those of neutron
stars, their frequencies are identified with test-particle
frequencies (Stella, Vietri \& Morsink 1999; Abramowicz et al.
2003). Most parts of an accretion disk, except, significantly, the
transition regions at the inner boundary of the disk, are
characterized by Keplerian rotation rates to a very good
approximation. At any radius beyond the transition region, acoustic,
magneto-acoustic and viscous corrections to the test-particle
Keplerian orbital frequency are negligible. In the outer disk regime
for disks around neutron stars and white dwarfs (and far enough from
a central black hole), test-particle frequencies for radial and
vertical perturbations of the orbit are degenerate with the
Keplerian orbital frequency. The degeneracy is removed if the
effects of general relativity are important. A Newtonian field of
tidal or higher multi-pole structure would also lead to a split in
the degeneracy. However, in Newtonian gravity, distortions of
stellar shape even for the most rapidly rotating neutron stars will
not introduce a significant level of non-degeneracy between the
frequencies of test-particle oscillations that is comparable to the
difference between the observed QPO frequency bands. Models
employing the Keplerian frequency have been relatively successful in
interpreting QPO frequency correlations. The Kepler (test particle)
frequency interpretation has also been employed to place constraints
on the masses and spins of compact objects. Empirically, predictions
based on test-particle frequencies applied to QPO frequency
correlations deviate from observations at a level of about 10\%.

The Keplerian frequency represents the basic imprint of rotation on
all the dynamical responses of the disk. Albeit simple, the
identification of QPOs with test-particle frequencies has a number
of shortcomings. First, precisely in a boundary layer where QPOs are
expected to be excited, the dynamical frequencies of oscillations
are not the test-particle frequencies. In the boundary layer,
viscous and magnetic forces lead to deviation of orbital frequencies
from Keplerian test-particle frequencies (see, e.g., Erkut \& Alpar
2004). The resulting band of non-Keplerian rotation frequencies
entail viscous and acoustic response and couplings between adjacent
rings of the disk fluid. Hydrodynamic effects are therefore
essential for an understanding of the frequency bands characterizing
the boundary layer. In particular, the degeneracy of test-particle
frequencies in the non-relativistic regime is lifted in the
hydrodynamic boundary layer, where the radial epicyclic frequency is
in fact larger than the orbital frequency. This simple observation
(Alpar \& Psaltis 2005) has important consequences for the
interpretation of kHz QPO frequencies, in particular for the
constraints on the neutron star mass-radius relation derived from
kHz QPOs.

Second, test-particle frequencies do not distinguish between
azimuthal sidebands. While kinematic models of QPOs like the
beat-frequency model involve one specific band of frequencies, say
$\omega $, a large number of sidebands, of frequencies $\omega
_{m}\cong \omega -m\Omega $, where $\Omega $ is the orbital
frequency, are allowed by azimuthal symmetry. Arguments as to why
only one or two QPO bands are excited must involve choices imposed
by the symmetries of the interaction between the magnetosphere and
the accretion disk boundary, leading to resonances with particular
frequencies at one or more radial regions in the disk. Without
resonances, test-particle frequency spectra present no distinction
between the modes. The realistic hydrodynamic modes may, in some
parameter ranges, distinguish between both fundamental modes and all
their azimuthal sidebands through the different growth or decay
rates of these modes, \emph{even in the case of free oscillations}.
A reduction of the number of relevant (easily excitable) modes of
free oscillations is certainly an important task for an
understanding of accretion disks around neutron stars, white dwarfs
and black holes. In the case of black holes, this classification of
free hydrodynamic modes in terms of their growth rates is even more
important as resonant excitation by the black hole is not available.

In this paper, we study the global modes of free oscillations at
some position $r$ in the inner disk-boundary or transition region by
analyzing the perturbed dynamical equations of a hydrodynamic disk,
including pressure gradients, viscous and magnetic stresses. We
consider the case of a disk around a neutron star with a
magnetosphere. The neutron star is taken to be a \emph{slow
rotator}: the star's rotation rate is less than the value of the
Keplerian rotation rate at the inner radius of the disk; $\Omega
_{\ast }<\Omega _{\mathrm{K}}(r_{\mathrm{in}})$. The actual rotation
rate of the disk is set by the boundary condition $\Omega
(r_{\mathrm{in}})=\Omega _{\ast }$. Thus in the steady state
solutions for the disk, the rotation rate at the inner edge of the
disk and in the transition region beyond is sub-Keplerian. The
innermost disk radius $r_{\mathrm{in}}$ determines the disk
magnetosphere interface which may be subject to magnetic
Rayleigh-Taylor or interchange instabilities. This instability
operates mainly at the disk-magnetosphere interface because of the
sharp density
contrast across the radius $r=r_{\mathrm{in}}$. The inner disk region for $r\gtrsim r_{\mathrm{%
in}}$ consists of a non-Keplerian boundary region that joins the
outer Keplerian disk with a continuous density distribution. As we
will see in \S\ \ref{ehdfom}, the surface density profile throughout
the non-Keplerian boundary region increases with decreasing radii.
Thus, our boundary region is not subject to interchange-like
instabilities.

The free oscillation modes we explore by perturbing a steady state solution
embody information about the stability of the disk through their growth or
decay rates. Growing modes will provide a clue as to the origin of the QPO
frequency bands, which could be excited by resonant forcing of the disk by
time dependent interactions with the star's magnetosphere. We find, as
expected, that growth or decay rates are determined by the dynamical effects
of viscosity, with an additional dependence on the sound speed and the
density and rotation-rate profiles in the non-Keplerian boundary region. For
some boundary region parameters, not all fundamental modes of free
oscillations and not all azimuthal sidebands grow, and among the growing
modes the growth rates differ. For other choices of boundary region
parameters the growth rates of all sidebands are similar. The frequencies of
the modes differ from test-particle frequencies by amounts that can be
several times larger than the corresponding QPO width. Most importantly, for
a reasonable set of accretion disk parameters, we can show that only a
certain few of the hydrodynamic free oscillation modes will grow, and that
the frequency bands of these oscillations can be associated with the
observed frequency bands. The nature of free oscillation modes in the
boundary region depends on whether the neutron star is rotating at a rate
slower or faster than the rotation rates prevailing in the inner boundary of
the disk. In this paper, we will consider the more common and
straightforward case of \emph{slow rotators}, the case when the neutron star
rotation rate $\Omega _{\ast }$ is less than $\Omega _{\mathrm{K}}(r_{%
\mathrm{in}})$, the Kepler rotation rate at some representative inner disk
radius $r_{\mathrm{in}}$. An investigation of \emph{forced} resonant
excitation of these prevalent modes, and the comparison and association with
observed QPO bands will follow in a subsequent paper.

\S\ 2 lays out the basic assumptions and equations, \S\ 3 displays
the mode analysis for the free oscillations, including a discussion
of hydrodynamic effects, frequency bands, and growth rates. In \S\
4, we discuss the results and present our conclusions.

\section{Basic Equations and Assumptions\label{beaa}}

We consider the excitation of oscillations in a geometrically thin
disk in vertical hydrostatic equilibrium. We write the continuity
and the radial and angular momentum equations in cylindrical
coordinates, integrated vertically over the disk thickness, as
\begin{equation}
\frac{\partial \Sigma }{\partial t}+\frac{1}{r}\frac{\partial }{\partial r}%
(r\Sigma v_{r})+\frac{1}{r}\frac{\partial }{\partial \phi }(\Sigma v_{\phi
})=0,  \label{avcnt}
\end{equation}
\begin{equation}
\frac{\partial v_{r}}{\partial t}+v_{r}\frac{\partial v_{r}}{\partial r}+%
\frac{v_{\phi }}{r}\frac{\partial v_{r}}{\partial \phi }-\frac{v_{\phi }^{2}%
}{r}=-\left( \frac{\partial \Phi }{\partial r}\right) _{z=H}-\frac{1}{\Sigma
}\frac{\partial \Pi }{\partial r}+\left\langle F_{r}\right\rangle ,
\label{avrm}
\end{equation}
and
\begin{equation}
\frac{\partial v_{\phi }}{\partial t}+v_{r}\frac{\partial v_{\phi }}{%
\partial r}+\frac{v_{\phi }}{r}\frac{\partial v_{\phi }}{\partial \phi }%
+v_{r}\frac{v_{\phi }}{r}=-\frac{1}{\Sigma r}\frac{\partial \Pi }{\partial
\phi }+\left\langle F_{\phi }\right\rangle .  \label{avazm}
\end{equation}
Here, $H$ is the half-thickness of the disk, $v_{r}$ and $v_{\phi }$
are the radial and azimuthal components of the velocity field in the
disk, $\Phi $ is the gravitational potential,
\begin{equation}
\Sigma \equiv \int_{-H}^{H}\rho dz  \label{smd}
\end{equation}
is the surface mass density,
\begin{equation}
\Pi \equiv \int_{-H}^{H}Pdz  \label{vip}
\end{equation}
is the vertically integrated thermal pressure, $\rho $ is the mass density, $%
P$ is the sum of the gas and radiation pressures, and
\begin{equation}
\left\langle F_{j}\right\rangle \equiv \frac{1}{\Sigma }\int_{-H}^{H}F_{j}%
\rho dz  \label{vamv}
\end{equation}
are the vertically averaged sums of the magnetic and viscous forces per unit
mass, and the subscript $j$ stands for the $r$ or $\phi$ component of the
force.

We only consider the $\phi $-component of the viscous force as the
dominant shear stress between adjacent layers, because of our
assumption that the disk is geometrically thin (Shakura \& Sunyaev
1973). We write the sum of the viscous and large-scale magnetic
forces (per unit mass) in the radial and azimuthal directions as
$F_{r}=F_{r}^{\mathrm{mag}}$ and $F_{\phi }=F_{\phi
}^{\mathrm{vis}}+F_{\phi }^{\mathrm{mag}}$, respectively.

The steady, axisymmetric, equilibrium state obeys
\begin{equation}
-2\pi r\Sigma _{0}v_{r0}=\dot{M},  \label{eqcnt}
\end{equation}
\begin{equation}
v_{r0}\frac{dv_{r0}}{dr}-\Omega ^{2}r=-\left( \frac{\partial \Phi }{\partial
r}\right) _{z=H}-\frac{1}{\Sigma _{0}}\frac{d\Pi _{0}}{dr}+\left\langle
F_{r}^{\mathrm{mag}}\right\rangle _{0},  \label{req}
\end{equation}
and
\begin{equation}
\frac{\kappa ^{2}}{2\Omega }v_{r0}=\left\langle F_{\phi }^{\mathrm{vis}%
}\right\rangle _{0} +\left\langle F_{\phi }^{\mathrm{mag}}\right\rangle _{0},
\label{azeq}
\end{equation}
where $\dot{M}$ is the constant mass-inflow rate, $v_{r0}<0$, $v_{\phi
0}=\Omega r$, $\Omega $ is the angular velocity of the disk plasma, and $%
\kappa \equiv \lbrack 2\Omega (2\Omega +rd\Omega /dr)]^{1/2}$ is the radial
epicyclic frequency.

We introduce perturbations to this axisymmetric fluid distribution
and use the subscripts \textquotedblleft 0\textquotedblright\ and
\textquotedblleft 1\textquotedblright\ to denote the equilibrium and
perturbed states,
respectively. The linearized perturbation equations follow from equations (%
\ref{avcnt})--(\ref{avazm}) as
\begin{equation}
\frac{d\Sigma _{1}}{d\tau }+\frac{1}{r}\frac{\partial }{\partial r}\left[
r(\Sigma _{0}v_{r1}+\Sigma _{1}v_{r0})\right] +\frac{1}{r}\frac{\partial }{%
\partial \phi }(\Sigma _{0}v_{\phi 1})=0,  \label{lcnt}
\end{equation}%
\begin{eqnarray}
\frac{dv_{r1}}{d\tau }-2\Omega v_{\phi 1}+\frac{\partial }{\partial r}%
(v_{r0}v_{r1}) &=&\frac{1}{\Sigma _{0}}\left( \frac{\Sigma _{1}}{\Sigma _{0}}%
\frac{d\Pi _{0}}{dr}-\frac{\partial \Pi _{1}}{\partial r}\right)   \notag \\
&&+\left\langle F_{r}^{\mathrm{mag}}\right\rangle _{1},  \label{lrd}
\end{eqnarray}%
and
\begin{eqnarray}
\frac{dv_{\phi 1}}{d\tau }+\frac{\kappa ^{2}}{2\Omega }v_{r1}+\frac{v_{r0}}{r%
}v_{\phi 1}+v_{r0}\frac{\partial v_{\phi 1}}{\partial r} &=&-\frac{1}{\Sigma
_{0}r}\frac{\partial \Pi _{1}}{\partial \phi }+\left\langle F_{\phi }^{%
\mathrm{vis}}\right\rangle _{1}  \notag \\
&&+\left\langle F_{\phi }^{\mathrm{mag}}\right\rangle _{1},  \label{liaz}
\end{eqnarray}%
where $d/d\tau \equiv \partial /\partial t+\Omega \partial /\partial \phi $
is the Lagrangian derivative following the motion of the fluid in the
azimuthal direction.

In the following, we study perturbations varying on timescales
shorter than the thermal timescale and thus expect that there is no
energy exchange between adjacent fluid elements. In this case, the
fluid elements respond adiabatically to density and pressure
perturbations and, at the same time, the disk thickness $H$, which
is determined by vertical hydrostatic equilibrium, responds to the
change of pressure. In general, we can write (Stehle \& Spruit 1999)
\begin{equation}
\Pi _{1}=\Sigma _{1}\left( \frac{\Pi _{0}}{\Sigma _{0}}\right) \left( \Gamma
+\frac{\partial \ln H}{\partial \ln \rho }\right) \;,  \label{dnsprt}
\end{equation}
where $\Gamma $ is the index of a polytropic equation of state. The terms in
the last parenthesis is of order unity. Because we will not be calculating
explicitly changes in the vertical equilibrium during the oscillations, we
will set the last term to unity. Therefore, we will be studying effectively
isothermal modes.

We consider long wavelength perturbations to analyze the lowest
order normal modes extending globally over the radial distance scale
$r_{\mathrm{out}}$ of the disk. For the stability of such global
modes the effect of cylindrical geometry and of the structure of the
unperturbed disk state must be taken into account. We perform a
local mode analysis. Our approach differs from the usual local mode
analysis of short-wavelength perturbations of radial dependence
$e^{ikr}$ in a WKB approximation. Consideration of modes with radial
wavenumbers would be appropriate for the analysis of local disk
modes in the short wavelength regime. Instead, we study the
stability of the global disk modes at each particular radius $r$ in
the boundary layer. In this region, the unperturbed disk quantities
such as the surface density $\Sigma _{0}$ and the
rotation rate $\Omega $ change on a length scale $(\Delta r)_{%
\mathrm{BL}}<r_{\mathrm{in}}$. Since the variation of the global
long wavelength perturbations has a much longer range
$r_{\mathrm{out}}\gg (\Delta r)_{\mathrm{BL}}$ in the radial
direction, we write equations~(\ref{lcnt})--(\ref{liaz}) at any
particular radius $r$ in dimensionless form, neglecting the
$r$-dependence of the perturbations:
\begin{equation}
\frac{d\sigma }{d\tau }+(1+\beta )\Omega _{s}u_{r}+\beta \Omega _{\nu
}\sigma +\Omega _{s}\frac{\partial u_{\phi }}{\partial \phi }=0,
\label{ctlc}
\end{equation}%
\begin{equation}
\frac{du_{r}}{d\tau }-2\Omega u_{\phi }+(1+\beta )\Omega _{\nu }u_{r}=\beta
\Omega _{s}\sigma +f_{r}^{\mathrm{mag}},  \label{rmlc}
\end{equation}%
\begin{equation}
\frac{du_{\phi }}{d\tau }+\frac{\kappa ^{2}}{2\Omega }u_{r}-\Omega _{\nu
}u_{\phi }=-\Omega _{s}\frac{\partial \sigma }{\partial \phi }+f_{\phi }^{%
\mathrm{vis}}+f_{\phi }^{\mathrm{mag}}.  \label{azlc}
\end{equation}%
Here, $\sigma \equiv \Sigma _{1}/\Sigma _{0}$ is the dimensionless density
perturbation, $u_{r}\equiv v_{r1}/c_{s}$ and $u_{\phi }\equiv v_{\phi
1}/c_{s}$ are the dimensionless velocity perturbations, $f_{r,\phi }\equiv
\left\langle F_{r,\phi }\right\rangle _{1}/c_{s}$ are the force
perturbations in units of frequency, and $\Omega _{\nu }\equiv -v_{r0}/r$
and $\Omega _{s}\equiv c_{s}/r$ are the typical frequencies associated with
the radial accretion and the effective sound speed $c_{s}=(\Pi _{0}/\Sigma
_{0})^{1/2}$, respectively. The parameter
\begin{equation}
\beta \equiv \left( \frac{d\ln \Sigma _{0}}{d\ln r}\right) _{r}  \label{bet}
\end{equation}%
represents the surface mass density profile at a particular radius
$r$ in the inner disk. Note that $\beta $ crucially represents the
structure of the unperturbed disk. This differs from the usual local
mode analysis in which the radial variation of the unperturbed
background quantities such as $\Sigma _{0}$ is neglected while the
radial derivatives of the perturbations are taken into account in
terms of their radial wavenumbers.

We look for solutions of the above linear set of equations of the form%
\begin{equation}
q(\phi ,t)=\sum\limits_{m}\int q_{m}(\omega )e^{i(m\phi -\omega t)}d\omega ,
\label{qnt}
\end{equation}
where the integer $m\geq 1$ represents non-axisymmetric modes
corresponding to the prograde motion of perturbations in the
azimuthal direction. The Fourier decomposition of equations
(\ref{ctlc})--(\ref{azlc}) yields
\begin{equation}
\left[ i(m\Omega -\omega )+\beta \Omega _{\nu }\right] \sigma _{m}+(1+\beta
)\Omega _{s}u_{r,m}+im\Omega _{s}u_{\phi ,m}=0,  \label{fdct}
\end{equation}
\begin{equation}
-\beta \Omega _{s}\sigma _{m}+\left[ i(m\Omega -\omega ) +(1+\beta
)\Omega_{\nu }\right] u_{r,m}-2\Omega u_{\phi ,m}=f_{r,m}^{\mathrm{mag}},
\label{fdrm}
\end{equation}
\begin{equation}
im\Omega _{s}\sigma _{m}+\frac{\kappa ^{2}}{2\Omega }u_{r,m}+\left[
i(m\Omega -\omega )-\Omega _{\nu }\right] u_{\phi ,m}=f_{\phi ,m}^{\mathrm{%
vis}} +f_{\phi ,m}^{\mathrm{mag}}.  \label{fdaz}
\end{equation}
To proceed further in the linear mode analysis, we need to specify
the force perturbations $f_{r,m}$ and $f_{\phi ,m}$ in equations
(\ref{fdrm}) and (\ref{fdaz}).

In this paper, we discuss the free oscillation modes. In a subsequent paper,
we will study oscillations forced by external perturbations, as in the case
of a compact object with an oblique large-scale magnetic field.

\section{Global Free Oscillations in a Viscous Accretion Disk\label{feos}}

When there are no perturbations introduced by large-scale magnetic
fields, i.e., when $f_{r,m}^{\mathrm{mag}}=f_{\phi
,m}^{\mathrm{mag}}=0$, the only non-negligible force perturbation is
due to the kinematic viscosity $\nu $. In the absence of an analytic
model for the effective kinematic viscosity in a turbulent shearing
flow, we adopt a simple damping force prescription, i.e., $F_{\phi
}^{\mathrm{vis}}=-v_{\phi }/t_{\nu }$, where $t_{\nu }$ is the
timescale for viscous accretion. In order for our prescription to
yield the order-of-magnitude estimate for the viscous timescale,
$t_{\nu }\sim r^{2}/\nu \sim r/\left\vert v_{r0}\right\vert $, we
write
\begin{equation}
\left\langle F_{\phi }^{\mathrm{vis}}\right\rangle =\gamma _{\nu }\frac{%
v_{r0}v_{\phi }}{r},  \label{vscf}
\end{equation}
where $\gamma _{\nu }$ is a dimensionless factor. Note that, for this
prescription of the viscous force to satisfy equation (\ref{azeq}) for the
steady equilibrium state, it is necessary that
\begin{equation}
\gamma _{\nu }=\frac{\kappa ^{2}}{2\Omega ^{2}}-\frac{\langle F_{\phi }^{%
\mathrm{mag}}\rangle _{0}}{v_{r0}\Omega }.  \label{gamnu}
\end{equation}
To linear order in the perturbed quantities, equation (\ref{vscf}) yields $%
\left\langle F_{\phi }\right\rangle _{1}=-\gamma _{\nu }\Omega _{\nu
}v_{\phi 1}$. The zeroth-order magnetic force enters the perturbation
equations only through the parameter $\gamma _{\nu }$ as given in equation (%
\ref{gamnu}).

For the case of interest, equations (\ref{fdct})--(\ref{fdaz}) become
\begin{equation}
\left[ i(m\Omega -\omega )+\beta \Omega _{\nu }\right] \sigma _{m}+
(1+\beta) \Omega _{s}u_{r,m}+im\Omega _{s}u_{\phi ,m}=0,  \label{hct}
\end{equation}
\begin{equation}
-\beta \Omega _{s}\sigma _{m}+\left[ i(m\Omega -\omega )+(1+\beta) \Omega
_{\nu }\right] u_{r,m}-2\Omega u_{\phi ,m}=0,  \label{hrm}
\end{equation}
\begin{equation}
im\Omega _{s}\sigma _{m}+\frac{\kappa ^{2}}{2\Omega }u_{r,m}+\left[
i(m\Omega -\omega )-(1-\gamma _{\nu })\Omega _{\nu }\right] u_{\phi ,m}=0,
\label{azh}
\end{equation}
yielding a dispersion relation of third order in $\omega $. We provide the
general expressions for the dispersion relation and the three complex
solutions in the Appendix.

For illustrative purposes, we write the expressions for the three
frequencies to leading order in $\Omega _{s}/\Omega $, i.e., for
small hydrodynamic corrections. The solutions of astrophysical
interest may actually be found in situations beyond this approximate
regime, as we shall discuss in \S\ 3.2. However, the approximate
solutions shed light on the basic parameters that determine the
frequencies and growth rates of the modes. In the limit of small
hydrodynamic corrections, i.e. $\varepsilon _{1} $, $\varepsilon
_{2}$, and $\varepsilon _{3}\ll 1$ (see Appendix), the axisymmetric
modes ($m=0$) have the frequencies and growth rates
\begin{eqnarray}
\mathrm{Re}\left[ \omega _{1,2}^{(0)}\right]  &=&\pm \kappa \pm \beta \left(
1+\beta \right) \left( \frac{\Omega _{s}}{2\kappa }\right) \Omega _{s}
\notag \\
&&\pm \left[ 3\left( \gamma _{\nu }-\beta -1\right) -\left( \gamma _{\nu
}-\beta \right) ^{2}\right] \left( \frac{\Omega _{\nu }}{6\kappa }\right)
\Omega _{\nu },  \label{ob0r}
\end{eqnarray}%
\begin{equation}
\mathrm{Im}\left[ \omega _{1,2}^{(0)}\right] =-\left( \frac{\gamma _{\nu
}+\beta }{2}\right) \Omega _{\nu },  \label{ob0i}
\end{equation}%
\begin{equation}
\mathrm{Re}\left[ \omega _{3}^{(0)}\right] =0,  \label{ou0r}
\end{equation}%
and
\begin{equation}
\mathrm{Im}\left[ \omega _{3}^{(0)}\right] =-\beta \Omega _{\nu }.
\label{ou0i}
\end{equation}%
The corresponding expressions for the non-axisymmetric modes, with positive
or negative integer values of the azimuthal wavenumber $m$ are
\begin{eqnarray}
\mathrm{Re}\left[ \omega _{1,2}^{(m)}\right]  &=&m\Omega \pm \kappa \pm %
\left[ m^{2}+\beta \left( 1+\beta \right) \right] \left( \frac{\Omega _{s}}{%
2\kappa }\right) \Omega _{s}  \notag \\
&&\pm \left[ 3\left( \gamma _{\nu }-\beta -1\right) -\left( \gamma _{\nu
}-\beta \right) ^{2}\right] \left( \frac{\Omega _{\nu }}{6\kappa }\right)
\Omega _{\nu },  \label{obmr}
\end{eqnarray}%
\begin{equation}
\mathrm{Im}\left[ \omega _{1,2}^{(m)}\right] =-\left( \frac{\gamma _{\nu
}+2\beta }{3}\right) \Omega _{\nu },  \label{obmi}
\end{equation}%
\begin{equation}
\mathrm{Re}\left[ \omega _{3}^{(m)}\right] =m\Omega ,  \label{oumr}
\end{equation}%
and
\begin{equation}
\mathrm{Im}\left[ \omega _{3}^{(m)}\right] =-\left( \frac{\gamma _{\nu
}+2\beta }{3}\right) \Omega _{\nu }.  \label{oumi}
\end{equation}%
Modes whose real frequencies differ only by a sign are of course identical.
We note that in these approximate expressions to leading order in $\Omega
_{s}$ the imaginary parts have no dependence on $m$. This means that the
fundamental modes and the infinite sequence of sidebands shifted in
frequency by $\pm |m|\Omega $ from the fundamental modes of interest, all
have the same growth or decay rates in the limit of small hydrodynamic
corrections, i.e. $\Omega _{\nu }\ll \Omega $ and $\Omega _{s}\ll \Omega $.
Any distinction between the free oscillation modes arises only in a regime
in which acoustic and viscous hydrodynamic effects are important.

It is possible to check the consistency of our stability analysis
using equations (\ref{ob0i}), (\ref{ou0i}), (\ref{obmi}), and
(\ref{oumi}) in the limit of non-magnetic local modes for which the
radial gradient in the background surface density at any given
location is negligible, that is, $d\ln \Sigma _{0}/d\ln r\ll 1$.
Note that none of the modes grows if the local
gradient of the surface density is neglected $(\beta =0)$ since $%
\gamma _{\nu }>0$ for $\langle F_{\phi }^{\mathrm{mag}}\rangle
_{0}=0$. There is no instability for local hydrodynamic modes in the
long wavelength limit. The global modes grow in the inner disk where
the radial gradient of $\Sigma _{0}$ cannot be neglected $(\beta
\neq 0).$

\subsection{Hydrodynamic Corrections to Test-particle Frequencies\label%
{hctpf}}

In the ideal case of $\Omega _{\nu }=\Omega _{s}=0$, the three
characteristic frequencies coincide with appropriate combinations of
test-particle frequencies in the disk, i.e., $\omega
_{1}^{(m)}\simeq \kappa +|m|\Omega $, $\omega _{2}^{(m)}\simeq
\kappa -|m|\Omega $, and $\omega _{3}^{(m)}\simeq |m|\Omega $. This
justifies, to zeroth order, the identification of the observed QPO
frequencies with frequencies of test particles in kinematic models
of the QPOs. For example, in the sonic-point model (Miller et al.
1998), the higher kHz QPO would be the $\omega _{3}^{(1)}$ mode at
the sonic radius. In the relativistic precession model (Stella et
al.\ 1999), the two high-frequency QPOs would be the $\omega
_{3}^{(1)}$ and $\omega _{2}^{(1)}$ modes.

In the absence of external forcing, high-frequency modes in an
accretion disk can be excited only in the presence of viscosity.
This is a well-known result, discussed in detail by Kato (2001).
Because the disk cannot respond thermally at the high-frequencies of
interest here, viscosity can excite the modes only through its
dynamical effect (mechanism II in Kato 2001) and not through its
thermal effect (mechanism I in Kato 2001). Thus free oscillation
modes in the accretion disk modes can have positive growth rates only if $%
\Omega _{\nu }$ is not zero, i.e., through the effect of viscosity,
and, hence, only if the hydrodynamic contributions to their
frequencies are operative. Indeed, all models of hydrodynamic
accretion disk modes show that non-negligible hydrodynamic
corrections modify the test-particle frequencies of growing modes
(see, e.g., Wagoner 1999; Kato 2001; Psaltis \& Norman 2000).

We estimate the expected hydrodynamic corrections to the oscillatory
mode frequencies typically identified with high-frequency QPOs using
the scaling of the viscous and thermal frequencies in an alpha disk,
i.e.,
\begin{equation}
\Omega _{\nu }\simeq \alpha \left( \frac{H}{r}\right) ^{2} \left( \frac{%
\Omega_{\mathrm{K}}}{\Omega}\right) ^{2} \Omega  \label{onuad}
\end{equation}
and
\begin{equation}
\Omega _{s}\simeq \left( \frac{H}{r}\right) \left( \frac{\Omega_{\mathrm{K}}%
}{\Omega}\right) \Omega.  \label{ados}
\end{equation}
The fractional hydrodynamic correction to, e.g., the $\mathrm{Re}\left[
\omega _{2}^{(1)}\right]\simeq \kappa -\Omega $ mode is determined by the
acoustic response provided that $\alpha $ is not close to unity:
\begin{equation}
\frac{\delta \omega _{2}^{(1)}}{\omega _{2}^{(1)}}\simeq \frac{\Omega
_{s}^{2}}{\kappa (\kappa -\Omega )}\simeq \left( \frac{H}{r}\right)
^{2}\left( \frac{\kappa }{\Omega }\right) ^{-1} \left( \frac{\Omega_{\mathrm{%
K}}}{\Omega}\right) ^{2} \left\vert 1-\frac{\kappa }{\Omega }\right\vert
^{-1}.  \label{eq:corr}
\end{equation}

The growth rate of the same modes is $\simeq \Omega _{\nu }$ and hence the
quality factor $Q$ of the corresponding QPO will be at most
\begin{equation}
Q\simeq \frac{\kappa -\Omega }{\Omega _{\nu }} \simeq \alpha ^{-1}\left(
\frac{H}{r}\right) ^{-2} \left( \frac{\Omega_{\mathrm{K}}}{\Omega}\right)
^{-2} \left\vert 1-\frac{\kappa }{\Omega }\right\vert  \label{eq:Q}
\end{equation}
and the resulting width of the QPO will be $\simeq |\Omega -\kappa |/Q
\simeq \Omega _{\nu }$. Combining equations~(\ref{eq:corr}) and (\ref{eq:Q})
we obtain for the hydrodynamic correction to the mode frequency in units of
the width of the corresponding QPO the relation
\begin{equation}
\frac{\delta \omega _{2}^{(1)}}{\omega _{2}^{(1)}/Q}\simeq \frac{1}{\alpha }
\left( \frac{\Omega }{\kappa }\right).  \label{hydcrc}
\end{equation}
It is clear from this expression that, even for relatively large values of
the parameter $\alpha \simeq 0.1$, the QPOs will have frequencies that are
distinct from the corresponding test-particle frequencies by amounts that
are several times larger than the QPO widths.

\begin{figure}[h]
\epsscale{1.0} \plotone{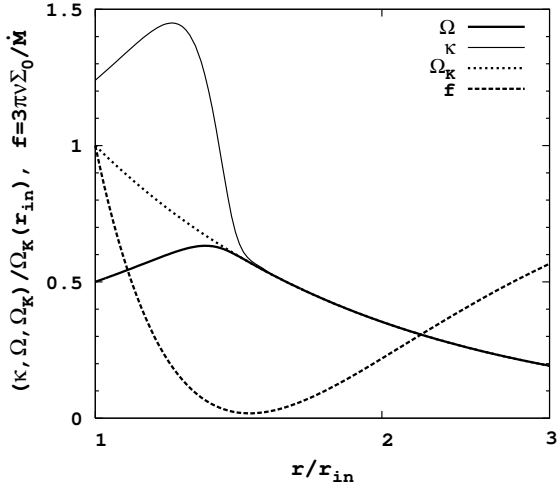} \caption{The radial profiles of the
orbital frequency $\Omega (r)$,
epicyclic frequency $\protect\kappa (r)$, Keplerian frequency $\Omega _{%
\mathrm{K}}(r)$, and the vertically integrated dynamical viscosity
$f(r)$ throughout the inner disk. This example is based on a typical
boundary region model from Erkut \& Alpar (2004). The frequencies
are given in units of $\Omega _{\mathrm{K}}(r_{\mathrm{in}})$. In
this particular example, the actual orbital frequency at the inner
disk boundary and the rotation rate of the neutron star are half the
Keplerian value at $r_{\mathrm{in}}$; $\Omega
(r_{\mathrm{in}})=\Omega _{*}= \Omega
_{\mathrm{K}}(r_{\mathrm{in}})/2$. \label{fig1}}
\end{figure}

\subsection{Excitation of Global Hydrodynamic Free Oscillation Modes\label%
{ehdfom}}

The growth rates of free oscillation modes are determined by a
number of parameters characterizing the boundary region in the inner
disk. In this region, the angular velocity of the disk matter makes
a transition from the Keplerian rotation to match the stellar
rotation rate. Our analysis depends on the local values of these
parameters at any particular radius $r$ within this transition or
boundary region. The main parameters are the ratio of the
radial epicyclic frequency $\kappa $ to the orbital angular frequency $%
\Omega $, the radial surface-density profile $\beta $ (equation \ref{bet}), $%
\Omega _{\nu }/\Omega _{s}$, the ratio of the radial drift velocity to the
sound speed, and $\Omega _{s}/\Omega $, the inverse timescale or the typical
frequency associated with the sound speed in units of the angular velocity.
In this section, we relate these parameters to the angular velocity profile $%
\Omega (r)$ and the dynamical viscosity $\nu \Sigma _{0} $ in the
accretion disk. We then study the local excitation of the modes in
the inner disk or boundary region where the hydrodynamic effects are
important.

The ratio $\kappa /\Omega $ of the epicyclic and orbital frequencies is
related to the orbital frequency profile $\Omega (r)$ through the relation
\begin{equation}
\frac{\kappa ^{2}}{\Omega ^{2}}=4\left( 1+\frac{1}{2}\frac{d\ln \Omega }{%
d\ln r}\right) .  \label{kappaOmega}
\end{equation}
In a non-Keplerian boundary-transition region of the accretion disk
around a neutron star that is a slow rotator, $\Omega(r)$ is less
than the Keplerian value $\Omega_{\mathrm{K}}(r)$. Proceeding from
the Keplerian outer disk through the transition region, $\Omega(r)$
goes through a maximum and then decreases to match the star's
rotation rate $\Omega _{\ast }$ at the inner
radius of the disk. The epicyclic frequency $\kappa$, degenerate with $%
\Omega_{\mathrm{K}}$ in the outer disk, increases through the
transition region. The ratio $\kappa /\Omega$ equals 2 at the radius
where $\Omega$ is maximum, has values between 1 and 2 in the outer
parts of the non-Keplerian transition region, and is larger than 2
in the inner parts. Fig.~1 shows the run of $\Omega(r)$ and $\kappa
(r)$ in a typical transition region. The numerical data for this
example are obtained from the model curve shown in the panel (a) of
Fig.~2 in Erkut \& Alpar (2004).

For a viscous accretion disk, we write the turbulent viscosity,
\begin{equation}
\nu =\alpha \frac{c_{s}^{2}}{\Omega },  \label{vscty}
\end{equation}
using the $\alpha $-prescription (Shakura \& Sunyaev 1973). The
angular velocity profile in a disk depends on the radial profile of
the vertically integrated dynamical viscosity,
\begin{equation}
\nu \Sigma _{0} =\frac{\dot{M}}{3\pi }f(r),  \label{vidv}
\end{equation}
and the appropriate boundary conditions. Here, $f(r)$ is a
dimensionless function for the dynamical viscosity. The run of $f$
as a function of the radial distance in the typical disk-transition
region is also shown in Fig.~1 (see Erkut \& Alpar 2004). The radial
profile of $f$ is independent of the particular prescription for
$\nu $. However, once we specify the kinematic viscosity as in
equation (\ref{vscty}), we can relate the surface
density profile (see equation \ref{bet}), using equations (\ref{vscty}) and (%
\ref{vidv}), to the orbital frequency profile and other parameters of the
boundary region through
\begin{equation}
\beta =\frac{d\ln f}{d\ln r}+\frac{d\ln \Omega }{d\ln r}-2\frac{d\ln c_{s}}{%
d\ln r} \equiv \beta _{0}-2\frac{d\ln c_{s}}{d\ln r}.  \label{beta}
\end{equation}
The first two terms, which we delineate as $\beta _{0}$ characterize each of
the boundary region models developed in Erkut \& Alpar (2004), with values
in the range $-15<\beta _{0}<3$.

Using equations (\ref{eqcnt}), (\ref{vscty}), and (\ref{vidv}), we
obtain the explicit dependence of the ratio of the radial drift
velocity and the sound speed on the disk-boundary region parameters
as
\begin{equation}
\frac{\Omega _{\nu }}{\Omega _{s}}=-\frac{v_{r0}}{c_{s}}=\frac{3}{2} \left(%
\frac{\alpha }{f} \right) \frac{\Omega _{s}}{\Omega }.  \label{onuovos}
\end{equation}
Note that this ratio, for the given values of $\alpha $ and $\Omega
_{s}/\Omega $, takes its maximum value for the minimum value of $f$. As $%
\Omega _{\nu }$, which determines the growth rates of the modes, depends on
the ratio $\Omega _{s}/\Omega $, the growth rates differ for sufficiently
large hydrodynamic corrections, i.e., $\Omega _{s}\lesssim \Omega $, and the
fastest growing modes can be identified. This is the regime of only slightly
supersonic azimuthal flow. In a magnetic boundary layer or transition zone
where $\Omega <\Omega _{\mathrm{K}}$, the speed of sound can be written as $%
c_{s}\simeq \Omega _{\mathrm{K}}\sqrt{Hr}$ whereas in the weakly
magnetized outer disk, $c_{s}\simeq \Omega _{\mathrm{K}}H$ (see
Erkut \& Alpar 2004). We therefore expect to find larger values for
$\Omega _{s}/ \Omega $ in a
non-Keplerian boundary region; $\Omega _{s}/ \Omega \simeq (\Omega _{\mathrm{%
K}}/ \Omega)(H/r)^{1/2}$. The larger the ratio $\Omega _{s}/\Omega $, the
stronger is the effect of hydrodynamic corrections on the growth rates of
the modes.

We explore the run of the frequencies and the growth rates of free
oscillations excited at a characteristic radius $r$ in the innermost
disk as the hydrodynamic parameter $\Omega _{s}/\Omega $ changes at
that radius. As a guideline, we have employed the model solutions of
Erkut \& Alpar (2004). These models qualitatively represent
conditions in a boundary-transition region. The key model parameters
can be easily translated into the
parameters in equations (\ref{kappaOmega}), (\ref{beta}), and (\ref{onuovos}%
). The zeroth-order magnetic force in equation (\ref{gamnu}) can also be
estimated from these model solutions with a range $0\leq \langle F_{\phi }^{%
\mathrm{mag}}\rangle _{0}/v_{r0}\Omega \lesssim 1$ in the inner boundary
region. The maximum value of this parameter corresponds to the radius where $%
\Omega $ is maximum. Near the innermost radius of the disk, the
azimuthal magnetic force vanishes as the relative shear between the
magnetosphere and the disk becomes negligible, i.e., $\Omega -\Omega
_{*}\simeq 0$.

To elucidate how the hydrodynamic parameters give different growth rates for
different modes, we investigate the excitation of free oscillations in the
typical boundary region shown in Fig.~1. To see the hydrodynamic effects on
the growth rates of the modes, we use the full eigenfrequency solutions for
axisymmetric and non-axisymmetric perturbations given in equations (\ref%
{afeo})--(\ref{thef}) and (\ref{fof})--(\ref{ftf}) (see Appendix). The panel
(a) of Fig.~2 shows the real parts of the complex mode frequencies at the
radius where $\Omega $ is maximum for the $m=0$ and $m=1$ cases. For the
boundary layer model chosen, the model parameters at this radius are $\kappa
/\Omega =2$, $\beta =-14$, $f=0.067$, and $\langle F_{\phi }^{\mathrm{mag}%
}\rangle _{0}/v_{r0}\Omega =1$. The panel (a) of Fig.~2 is obtained for $%
\alpha =0.1$. Although the real parts of the mode frequencies do not change
under a different value of the viscosity parameter, say $\alpha =0.01$ at
this radius, the imaginary parts do (see panel b of Fig.~2). The unstable
modes grow faster for $\alpha =0.1$ than they do for $\alpha =0.01$, as
shown in Fig.~2 (panel b). This is because the growth rates are primarily
determined by $\Omega _{\nu }/\Omega _{s}\propto \alpha /f$ (see equation %
\ref{onuovos}). In Fig.~3, we display the real (panel a) and imaginary
(panel b) parts of the mode frequencies at the innermost disk radius where $%
\Omega =\Omega _{*}$ for $m=0$ and $m=1$. The model parameters illustrated
in Fig.~3 are $\kappa /\Omega =2.5$, $\beta =-5$, $f=1$, and $\langle
F_{\phi }^{\mathrm{mag}}\rangle _{0}/v_{r0}\Omega =0$. Fig.~3 is plotted for
$\alpha =0.1$. Note that the growth rates, for the same value of the
viscosity parameter $\alpha $, are lower than those at the radius where $%
\Omega $ is maximum.

The analysis for the particular transition zone discussed above can also be
extended to other sample boundary regions with different model parameters.
Our conclusion that the fastest growing modes are excited near the radius $%
r=r_{0}$ where $\Omega $ is maximum remains valid for all
sub-Keplerian inner disks around slow rotators: The common property
of all boundary regions is that the parameters $|\beta |$ and
$\alpha /f$, which determine the growth rates of the unstable modes,
attain highest values near $r_{0}$ where $\kappa /\Omega \simeq 2$.
Note that the growing mode frequencies
significantly exceed test-particle frequencies in the boundary region for $%
\Omega _{s}/\Omega \lesssim 1$. In the limit of small hydrodynamic
corrections ($\Omega _{s}/\Omega \ll 1$), the mode frequencies converge to
the corresponding test-particle frequencies (see panel a of Fig.~2 and panel
a of Fig.~3) with negligible growth rates as expected (see panel b of Fig.~2
and panel b of Fig.~3).

The frequency branches that map to the test-particle frequencies $\omega
_{1}^{(0)} =\kappa $ and $\omega _{1,2}^{(1)} =\kappa \pm \Omega $ in the
limit of small hydrodynamic corrections, $\Omega _{s}/\Omega \ll 1$, have
positive and rather fast growth rates at $r\lesssim r_{0}$, where $\kappa
/\Omega \gtrsim 2$, when $\Omega _{s}/\Omega \lesssim 1$ (see panel a of
Fig.~2 and panel a of Fig.~3). We identify $\omega _{3}^{(0)} =0$ and $%
\omega _{3}^{(1)} \simeq \Omega $ as the quasi-stable or decaying mode
frequencies in the same range of $\Omega _{s}/\Omega $ as shown in Fig.~2
(panel b) and Fig.~3 (panel b). Note that while each of the growing branches
differs from test-particle frequencies as $\Omega _{s}/\Omega $ increases,
the difference frequency between consecutive bands is always $\Delta \omega
= \omega _{1}^{(1)}-\omega _{1}^{(0)}\lesssim \Omega $ or $\Delta \omega =
\omega _{1}^{(0)}-\omega _{2}^{(1)}\lesssim \Omega $ over a wide range of $%
\Omega _{s}/\Omega $ (see panel a of Fig.~2 and panel a of Fig.~3). As $%
\Omega \gtrsim \Omega _{*}$ in a boundary layer, we find $\Delta \omega
\simeq \Omega _{*}$ as observed for the relatively slow rotators among the
neutron-star LMXBs that exhibit kHz QPOs.

\section{Discussion and Conclusions}

We have studied the global hydrodynamic modes of long wavelength in
the boundary or transition region of viscous accretion disks as a
possible source of kHz QPOs in neutron-star LMXBs. The stability of
the eigenmodes strongly depends on the global disk structure imposed
by the boundary conditions as expected from the mode analysis in the
long wavelength regime. Our local treatment takes account of the
local effects of the global disk parameters on the excitation of
hydrodynamic free oscillations.

We find that the frequencies and growth rates of the modes are
mainly determined by global disk parameters such as $\kappa /\Omega
$, the ratio of the radial epicyclic frequency to the orbital
frequency, the radial surface-density profile $\beta $, $\Omega
_{\nu }/\Omega _{s}$, the ratio of the radial drift velocity to the
sound speed, and $\Omega _{s}/\Omega $, the
ratio of the sound speed to the azimuthal velocity. The local values of $%
\kappa /\Omega $ and $\beta $ directly follow from the global
solution for the rotational dynamics of a boundary-transition region
model. The parameters $\Omega _{\nu }/\Omega _{s}$ and $\Omega
_{s}/\Omega $ depend additionally on the particular viscosity
prescription and the details of the structure of accretion flow in
the inner disk.

The hydrodynamic effects on the frequencies and growth rates of the
modes are reflected by the parameters $\Omega _{\nu }/\Omega _{s}$
and $\Omega _{s}/\Omega $. We have found that the growth rates of
different modes for a given azimuthal wavenumber $m$ differ
significantly when the hydrodynamic corrections are sufficiently
large, i.e. $\Omega _{\nu}/\Omega _{s}\lesssim 1 $ and $\Omega
_{s}/\Omega \lesssim 1$. In this regime, the growing mode
frequencies significantly exceed test-particle frequencies for a
plausible range of $\Omega _{s}/\Omega $ at any particular radius
within the boundary region. This is because the effect of
hydrodynamic corrections on the growth rates of the modes is
expected to be strong in a magnetic boundary layer or transition
zone as discussed in \S\ 3.2. In the limit of small hydrodynamic
corrections, i.e. $\Omega _{\nu}/\Omega _{s}\ll 1$ and $\Omega
_{s}/\Omega \ll 1$, the eigenmodes have test-particle frequencies
with negligible growth rates. Our analysis shows that taking account
of hydrodynamic effects has an important outcome: the frequencies of
the growing modes gain higher values above test-particle frequencies
as their growth rates increase for larger hydrodynamic corrections.

Throughout the boundary region, we have identified the growing mode
frequencies by the eigenvalues that approach the test-particle frequencies $%
\omega _{1}^{(0)} =\kappa $, $\omega _{1}^{(1)} =\kappa +\Omega $, and $%
\omega _{2}^{(1)} =\kappa -\Omega $ in the limit of small hydrodynamic
corrections (see panel a of Fig.~2). Modes such as $\omega _{3}^{(0)} =0$
and $\omega _{3}^{(1)}\simeq \Omega $ are not excited for the range of $%
\Omega _{s}/\Omega $ used in Fig.~2 (panel b) and Fig.~3 (panel b).
The difference frequency between successive bands of growing modes
is $\Delta \omega \simeq \Omega _{*}$ as observed for the relatively
slowly rotating neutron stars in LMXBs that show kHz QPOs (\S\ 3.2).
In a boundary layer or transition region, the orbital frequency of
the disk matter is close to the rotation frequency of the star
($\Omega \sim \Omega _{*}$). As $\Delta \omega \simeq \Omega$, the
separation between consecutive mode frequencies is always related to
the spin frequency of the neutron star. This is the result of the
boundary condition imposed by the rotating magnetosphere on the
innermost disk. The modes of the test-particle frequencies $\omega
_{1}^{(0)}\cong \kappa $ and $\omega _{2}^{(1)}\cong \kappa -\Omega
$ were originally proposed by Alpar \& Psaltis (2005) to be
associated with the upper and lower kHz QPOs observed in
neutron-star LMXBs. The frequency separation between the higher and
lower-frequency kHz QPO peaks decreases by a few tens of Hz as both
QPO frequencies increase by hundreds of Hz in a range of $200-1200$
Hz per source in all observed sources, namely, Sco~X-1,
4U~$1608-52$, 4U~$1728-34$ and 4U~$1735-44$ (van der Klis 2000).
This pair of frequencies satisfies the observed behavior that the
difference frequency $\Delta \omega $ between two kHz QPOs decreases
as both frequencies increase. According to our present analysis, the
pairs of observed kHz QPO bands could be either $\omega
_{1}^{(1)}\simeq \kappa +\Omega $ and $\omega _{1}^{(0)}\simeq
\kappa $ or $\omega _{1}^{(0)}\simeq \kappa $ and $\omega
_{2}^{(1)}\simeq \kappa -\Omega $, respectively, since the
identification of the observed QPOs with $\omega _{1}^{(1)}\simeq
\kappa +\Omega $ and $\omega _{1}^{(0)}\simeq \kappa $ also meets
the observed condition that $\Delta \omega $ decreases when both QPO
frequencies increase. The growth rates of free oscillations produce
a wide spectrum of sidebands, $\omega _{1,2}^{(m)}\cong \kappa \pm
\vert m\vert \Omega $, some of which fall in the range of observed
power spectra. The reduction to only two of kHz QPO bands as
observed in neutron star sources must therefore be a product of
forced oscillations, resonances, and boundary conditions. The role
of forced oscillations and resonances in the excitation of
particular modes will be discussed in future work.

We find that the fastest growing modes with frequency branches
$\kappa $ and $\kappa \pm \Omega $ are excited near the disk radius
$r\lesssim r_{0}$, where the orbital frequency is maximum. In the
boundary layer, the loss of centrifugal support near $r_{0}$ leads
to some radial acceleration of the disk matter. The subsequent rise
in the radial drift velocity of the disk matter is accompanied by a
sudden drop in the surface density as the vertically integrated
dynamical viscosity is minimized (see \S\ 3.2). This picture is
common to all boundary layers or transition zones for which magnetic
braking is efficient (see Erkut \& Alpar 2004 and references
therein). Thus, independent of the boundary conditions and the
magnetic field configuration in the boundary region, there exists a
specific disk radius where the hydrodynamical background quantities
such as the surface density $\Sigma _{0}$ and the radial drift
velocity $v_{r0}$ change dramatically. The steepness in the change
of the surface density in the
radial direction is reflected through the parameter $\beta $ (see equation %
\ref{bet}). The sign and magnitude of $\beta $ (the surface density
profile in the disk) control whether the free oscillation modes will
grow or decay. We see that the modes of frequency bands $\kappa $
and $\kappa \pm \Omega $
can be excited with the largest growth rates only for the disk radii $%
r\lesssim r_{0}$, where $\beta $ is minimum and $\Omega $ is maximum
(see \S\ 3.2 for the range of $\beta $). The inverse timescale,
$\Omega _{\nu} $, associated with radial accretion also obtains its
highest value near the same radii yielding the maximum growth rates
in the accretion disk.

The boundary layers with sub-Keplerian rotation rates are usually
expected to be realized in the innermost regions of accretion disks
around \emph{slow rotators}. For a \emph{slow rotator}, the Kepler
rotation rate at the inner disk radius $\Omega
_{\mathrm{K}}(r_{\mathrm{in}})$ prevails over the stellar rotation
rate $\Omega _{*}$. As we have illustrated in \S\ 3.2, the
difference frequency between two consecutive bands of growing modes
in the boundary region is close to the spin frequency of the neutron star ($%
\Delta \omega \simeq \Omega _{*}$). This result agrees with the kHz QPO
observations of the relatively slow rotators with spin frequencies below $%
\sim 400$~Hz in neutron-star LMXBs. In these sources, for example,
in 4U~$1728-34$, the lower kHz QPO frequencies increase from 600 to
900 Hz and the upper kHz QPO frequencies increase from 950 to 1200
Hz, while the difference between the two observed kHz QPO frequency
bands decreases slightly, of the order of 10 Hz, while remaining
still commensurate with the spin frequency of the neutron star
(M\'{e}ndez \& van der Klis 1999). A second class of kHz QPO
sources, such as the sources KS~$1731-260$, Aql~X-1, 4U~$1636-53$,
which have spin frequencies above $\sim 400$~Hz is usually depicted
as \emph{fast rotators} (see Wijnands et al. 2003). In this second
class of sources, for example in 4U~$1636-53$, the lower kHz QPO
frequencies increase from 900 to 950 Hz and the upper kHz QPO
frequencies increase from 1150 to 1190 Hz, while the difference
between the two observed kHz QPO frequency bands again decreases by
amounts of the order of 10 Hz (see van der Klis 2000 and references
therein). The frequency separation between the kHz QPOs is close to
half the spin frequency of the neutron star if it is a \emph{fast
rotator}. The emergence of two seemingly different classes of
neutron-star LMXBs could be related to whether the rotation rate of
the neutron star is less or greater than that of the inner disk
matter. It is probable that $\Omega _{*}>\Omega
_{\mathrm{K}}(r_{\mathrm{in}})$ for the relatively fast rotating
neutron star sources for which the difference frequency between twin
QPO peaks is $\Delta \omega \simeq \Omega _{*}/2$. Recent work by
M\'{e}ndez \& Belloni (2007) actually suggests a more
continuous distribution between $\Delta \omega \simeq \Omega _{*}$ and $%
\Delta \omega \simeq \Omega _{*}/2$. The rotational dynamics of
accretion flow in the innermost regions of accretion disks around
these \emph{fast rotators} could be quite different from those of
the transition regions we consider here. We plan to address this
issue in a subsequent paper.

\acknowledgments
We acknowledge support from the Marie Curie FP6 Transfer of Knowledge
Project ASTRONS, MKTD-CT-2006-042722. M.\,H.\,E.\ was partially supported by
a T\"{U}B\.{I}TAK (The Scientific and Technical Research Council of Turkey)
postdoctoral fellowship. D.\,P.\ was partially supported by NASA grant
NAG-513374. M.\,A.\,A.\ acknowledges support from the Turkish Academy of
Sciences and the Sabanc{\i} University Astrophysics and Space Forum.

\appendix

\section{General Expressions}

The dispersion relation of the global force-free hydrodynamic modes
we consider in \S\ 3 is
\begin{equation}
(\omega -m\Omega )\left[(\omega -m\Omega )^{2}-\Omega _{a}^{2}\right]-\Omega
_{b}^{3} -i\left[ (\omega -m\Omega )^{2}\Omega _{c}+\Omega _{d}^{3}\right]
=0,  \label{dr}
\end{equation}
where
\begin{equation}
\Omega _{a}^{2} \equiv \kappa ^{2}+\left[ m^{2}+\beta (1+\beta )\right]%
\Omega_{s}^{2} +\left[ \beta (1+\beta )-(1+2\beta )(1-\gamma _{\nu })\right]%
\Omega _{\nu }^{2},  \label{oma}
\end{equation}
\begin{equation}
\Omega _{b}^{3}\equiv 2m(1+\beta )\Omega _{s}^{2}\Omega +m\beta \Omega
_{s}^{2}\left( \frac{\kappa ^{2}}{2\Omega }\right) ,  \label{omb}
\end{equation}
\begin{equation}
\Omega _{c}\equiv -(\gamma _{\nu }+2\beta )\Omega _{\nu },  \label{omc}
\end{equation}
and
\begin{equation}
\Omega _{d}^{3} \equiv \beta \Omega _{\nu }\kappa ^{2}-\beta (1+\beta
)(1-\gamma _{\nu })\Omega _{\nu }^{3} -(1+\beta )\left[ \beta (1-\gamma
_{\nu })-m^{2}\right] \Omega _{\nu }\Omega _{s}^{2}\;.  \label{omd}
\end{equation}

Note that the solution of equation (\ref{dr}), in the ideal case of
an inviscid $\Omega _{\nu }=0$, cold disk $\Omega _{s}\simeq 0$,
where the gas particles barely interact with each other, is given by
$\omega
_{1,2}^{(m)}\simeq m\Omega \pm \kappa $ and $\omega_{3}^{(m)}\simeq m\Omega $%
. In general, equation (\ref{dr}), for the realistic case of $\Omega _{\nu
}\neq 0$ and $\Omega _{s}\neq 0$, yields
\begin{equation}
\omega _{\mathrm{I}}^{(0)}=(\chi -\psi )\Omega _{d}+i\frac{\Omega _{c}}{3},
\label{afeo}
\end{equation}
\begin{equation}
\omega _{\mathrm{II}}^{(0)}=\frac{1}{2}(\psi -\chi )\Omega _{d}+i\frac{%
\Omega _{c}}{3}+i\frac{\sqrt{3}}{2}(\chi +\psi )\Omega _{d},  \label{ttef}
\end{equation}
and
\begin{equation}
\omega _{\mathrm{III}}^{(0)}=\frac{1}{2}(\psi -\chi )\Omega _{d}+i\frac{%
\Omega _{c}}{3}-i\frac{\sqrt{3}}{2}(\chi +\psi )\Omega _{d}  \label{thef}
\end{equation}
as the eigenfrequency solutions for axisymmetric perturbations ($m=0$),
where
\begin{equation}
\psi \equiv \frac{1}{9\chi }\left( \frac{\Omega _{c}}{\Omega _{d}}\right)
^{2}-\frac{1}{3\chi }\left( \frac{\Omega _{a}}{\Omega _{d}}\right) ^{2},
\label{cgmm}
\end{equation}
\begin{equation}
\chi \equiv \left( i\frac{\eta _{1}}{4}+\frac{1}{2}\sqrt{\eta _{2}-\eta _{1}}%
\right) ^{1/3},  \label{chi}
\end{equation}
\begin{equation}
\eta _{1}\equiv 2+\frac{2}{3}\left( \frac{\Omega _{a}^{2}\Omega _{c}}{\Omega
_{d}^{3}}\right) -\frac{4}{27}\left( \frac{\Omega _{c}}{\Omega _{d}}\right)
^{3},  \label{feta}
\end{equation}
and
\begin{equation}
\eta _{2}\equiv 1-\frac{4}{27}\left( \frac{\Omega _{a}}{\Omega _{d}}\right)
^{6}+\frac{1}{27}\left( \frac{\Omega _{a}^{4}\Omega _{c}^{2}}{\Omega
_{d}^{6} }\right).  \label{seta}
\end{equation}

For small hydrodynamic corrections, $\eta _{2}<\eta _{1}$, we obtain
\begin{equation}
\mathrm{Re}\left[ \omega _{1}^{(0)}\right] = \frac{\left( \sqrt{%
1+\varepsilon _{1}}+\varepsilon _{2}\right) ^{1/3}\Omega _{a}}{2} +\frac{%
\left(1-\varepsilon_{3}\right)\Omega _{a}}{2\left( \sqrt{1+\varepsilon _{1}}
+\varepsilon _{2}\right) ^{1/3}},  \label{aef}
\end{equation}
\begin{equation}
\mathrm{Im}\left[ \omega _{1}^{(0)}\right] =\frac{\left( \sqrt{1+\varepsilon
_{1}}+\varepsilon _{2}\right) ^{1/3}\Omega _{a}}{2\sqrt{3}} -\frac{\left(
1-\varepsilon _{3}\right) \Omega _{a}}{2\sqrt{3}\left( \sqrt{1+\varepsilon
_{1}}+\varepsilon _{2}\right) ^{1/3}}+\frac{\Omega _{c}}{3},  \label{iomo}
\end{equation}
\begin{equation}
\mathrm{Re}\left[ \omega _{2}^{(0)}\right] =-\,\mathrm{Re}\left[ \omega
_{1}^{(0)}\right] ,  \label{rotmz}
\end{equation}
\begin{equation}
\mathrm{Im}\left[ \omega _{2}^{(0)}\right] =\mathrm{Im}\left[ \omega
_{1}^{(0)}\right] ,  \label{iotmz}
\end{equation}
\begin{equation}
\mathrm{Re}\left[ \omega _{3}^{(0)}\right] =0,  \label{otrmz}
\end{equation}
\begin{equation}
\mathrm{Im}\left[ \omega _{3}^{(0)}\right] =\Omega _{c}-2\,\mathrm{Im}\left[
\omega _{1}^{(0)}\right]  \label{otimz}
\end{equation}
as the only non-degenerate eigenfrequency solutions to the dispersion
relation for axisymmetric perturbations, where
\begin{equation}
\varepsilon _{1}\equiv \frac{27}{4}\left( \frac{\Omega _{d}}{\Omega _{a}}%
\right) ^{6}+\frac{9\Omega _{c}\Omega _{d}^{3}}{2\Omega _{a}^{4}}-\frac{%
\Omega _{d}^{3}\Omega _{c}^{3}}{\Omega _{a}^{6}}-\frac{1}{4}\left( \frac{%
\Omega _{c}}{\Omega _{a}}\right) ^{2},  \label{epso}
\end{equation}
\begin{equation}
\varepsilon _{2}\equiv \frac{3\sqrt{3}}{2}\left( \frac{\Omega _{d}}{\Omega
_{a}}\right) ^{3}+\frac{\sqrt{3}\Omega _{c}}{2\Omega _{a}}-\frac{\sqrt{3}}{9}%
\left( \frac{\Omega _{c}}{\Omega _{a}}\right) ^{3},  \label{epst}
\end{equation}
and
\begin{equation}
\varepsilon _{3}\equiv \frac{1}{3}\left( \frac{\Omega _{c}}{\Omega _{a}}%
\right) ^{2}.  \label{epsh}
\end{equation}

The general solutions to the dispersion equation for $m\neq 0$ are
\begin{equation}
\omega _{\mathrm{I}}^{(m)}=m\Omega +(\xi +\Lambda )\Omega _{b}+i\frac{\Omega
_{c}}{3},  \label{fof}
\end{equation}
\begin{equation}
\omega _{\mathrm{II}}^{(m)}=m\Omega -\frac{1}{2}(\xi +\Lambda )\Omega _{b}+i%
\frac{\Omega _{c}}{3}+i\frac{\sqrt{3}}{2}(\xi -\Lambda )\Omega _{b},
\label{ff}
\end{equation}
and
\begin{equation}
\omega _{\mathrm{III}}^{(m)}=m\Omega -\frac{1}{2}(\xi +\Lambda )\Omega _{b}+i%
\frac{\Omega _{c}}{3}-i\frac{\sqrt{3}}{2}(\xi -\Lambda )\Omega _{b},
\label{ftf}
\end{equation}
where
\begin{equation}
\Lambda \equiv \frac{1}{3\xi }\left( \frac{\Omega _{a}}{\Omega _{b}}\right)
^{2}-\frac{1}{9\xi }\left( \frac{\Omega _{c}}{\Omega _{b}}\right) ^{2},
\label{lmbd}
\end{equation}
\begin{equation}
\xi \equiv \left( \frac{1}{2}+i\frac{\zeta _{1}}{4}+\frac{1}{2}\sqrt{\zeta
_{2}+i\zeta _{1}}\right) ^{1/3}\,  \label{ksi}
\end{equation}
\begin{equation}
\zeta _{1}\equiv \frac{2}{3}\left( \frac{\Omega _{a}^{2}\Omega _{c}}{\Omega
_{b}^{3}}\right) +2\left( \frac{\Omega _{d}}{\Omega _{b}}\right) ^{3}-\frac{4%
}{27}\left( \frac{\Omega _{c}}{\Omega _{b}}\right) ^{3},  \label{zto}
\end{equation}
and
\begin{equation}
\zeta _{2}\equiv 1-\frac{4}{27}\left( \frac{\Omega _{a}}{\Omega _{b}}\right)
^{6}+\frac{1}{27}\left( \frac{\Omega _{a}^{4}\Omega _{c}^{2}}{\Omega _{b}^{6}%
}\right) +\left( \frac{\Omega _{d}}{\Omega _{b}}\right) ^{6}-\zeta
_{1}\left( \frac{\Omega _{d}}{\Omega _{b}}\right) ^{3}.  \label{ztt}
\end{equation}

In the limit of small hydrodynamic corrections, $|\zeta _{1}|\,\ll |\zeta
_{2}|$, we find
\begin{equation}
\mathrm{Re}\left[ \omega _{1}^{(m)}\right] =m\Omega \pm \frac{\Omega
_{b}\Omega _{a}}{\left\vert \Omega _{b}\right\vert }\mp \left( \frac{\Omega
_{c}}{\Omega _{a}}\right) ^{2}\frac{\left\vert \Omega _{b}\right\vert \Omega
_{a}}{6\Omega _{b}},  \label{naor}
\end{equation}
\begin{equation}
\mathrm{Im}\left[ \omega _{1}^{(m)}\right] =\frac{\Omega _{c}}{3}\left[ 1\pm
\left( \frac{\Omega _{c}}{\Omega _{a}}\right) \frac{\left\vert \Omega
_{b}\right\vert }{2\sqrt{3}\Omega _{b}}\right] ,  \label{naoi}
\end{equation}
\begin{equation}
\mathrm{Re}\left[ \omega _{2}^{(m)}\right] =m\Omega \mp \frac{\Omega
_{b}\Omega _{a}}{\left\vert \Omega _{b}\right\vert }\pm \left( \frac{\Omega
_{c}}{\Omega _{a}}\right) ^{2}\frac{\left\vert \Omega _{b}\right\vert \Omega
_{a}}{6\Omega _{b}},  \label{natr}
\end{equation}
\begin{equation}
\mathrm{Im}\left[ \omega _{2}^{(m)}\right] =\mathrm{Im}\left[ \omega
_{1}^{(m)}\right] ,  \label{nati}
\end{equation}
\begin{equation}
\mathrm{Re}\left[ \omega _{3}^{(m)}\right] =m\Omega ,  \label{rnat}
\end{equation}
\begin{equation}
\mathrm{Im}\left[ \omega _{3}^{(m)}\right] =\Omega _{c}-2\,\mathrm{Im}\left[
\omega _{1}^{(m)}\right]  \label{inat}
\end{equation}
as the only non-degenerate eigenfrequency solutions to the dispersion
relation for non-axisymmetric perturbations.


\clearpage

\begin{figure}[tbp]
\epsscale{1.0} \plottwo{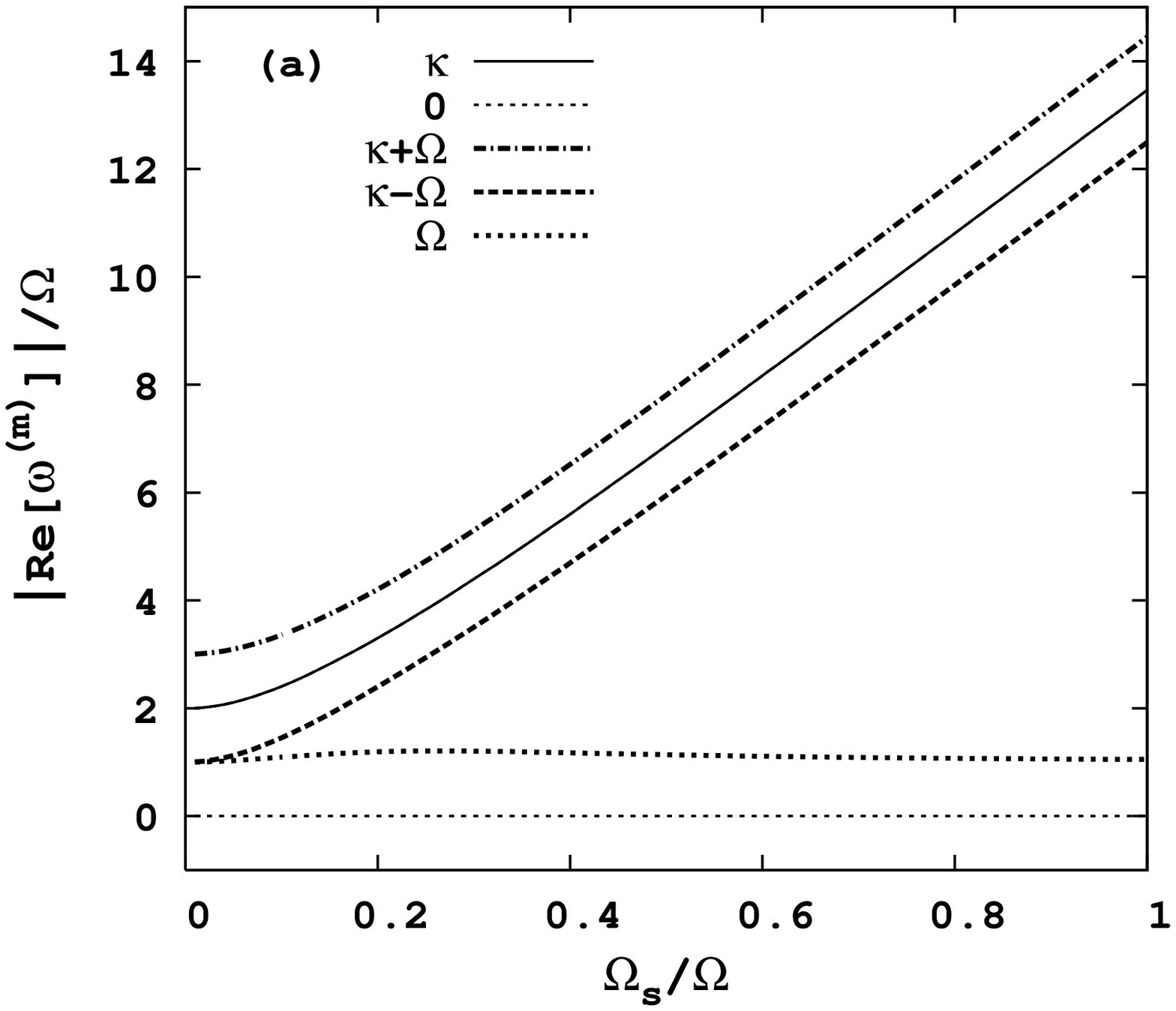}{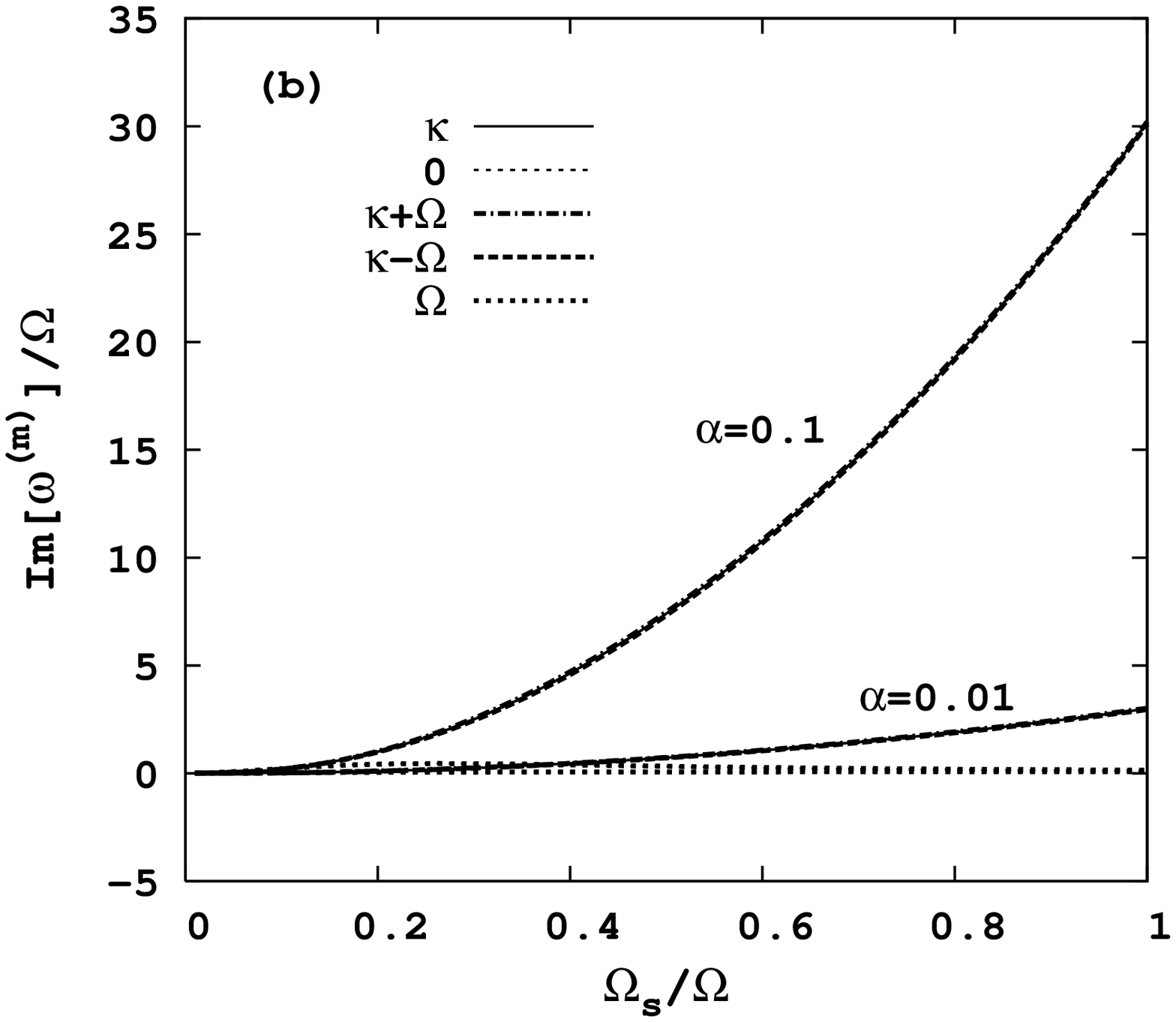} \caption{The real and
imaginary parts of the solutions to the dispersion relation for
axisymmetric ($m=0$) and non-axisymmetric ($m=1$) perturbations
excited at the radius where the orbital frequency $\Omega $ is
maximum. The
model parameters are $\protect\kappa /\Omega =2$, $\protect\beta =-14$, $%
f=0.067$, and $\langle F_{\protect\phi }^{\mathrm{mag}}\rangle
_{0}/v_{r0}\Omega =1$. The curves are labelled with the values of
the
frequencies in the test-particle limit. The $\protect\kappa $ and $\protect%
\kappa \pm \Omega $ modes have the same positive growth rates,
growing more rapidly for larger $\protect\alpha $. The
zero-frequency and $\Omega $
modes, indicated by the overlapping dotted curves, do not grow for any $%
\protect\alpha $. \label{fig2}}
\end{figure}


\begin{figure}[tbp]
\epsscale{1.0} \plottwo{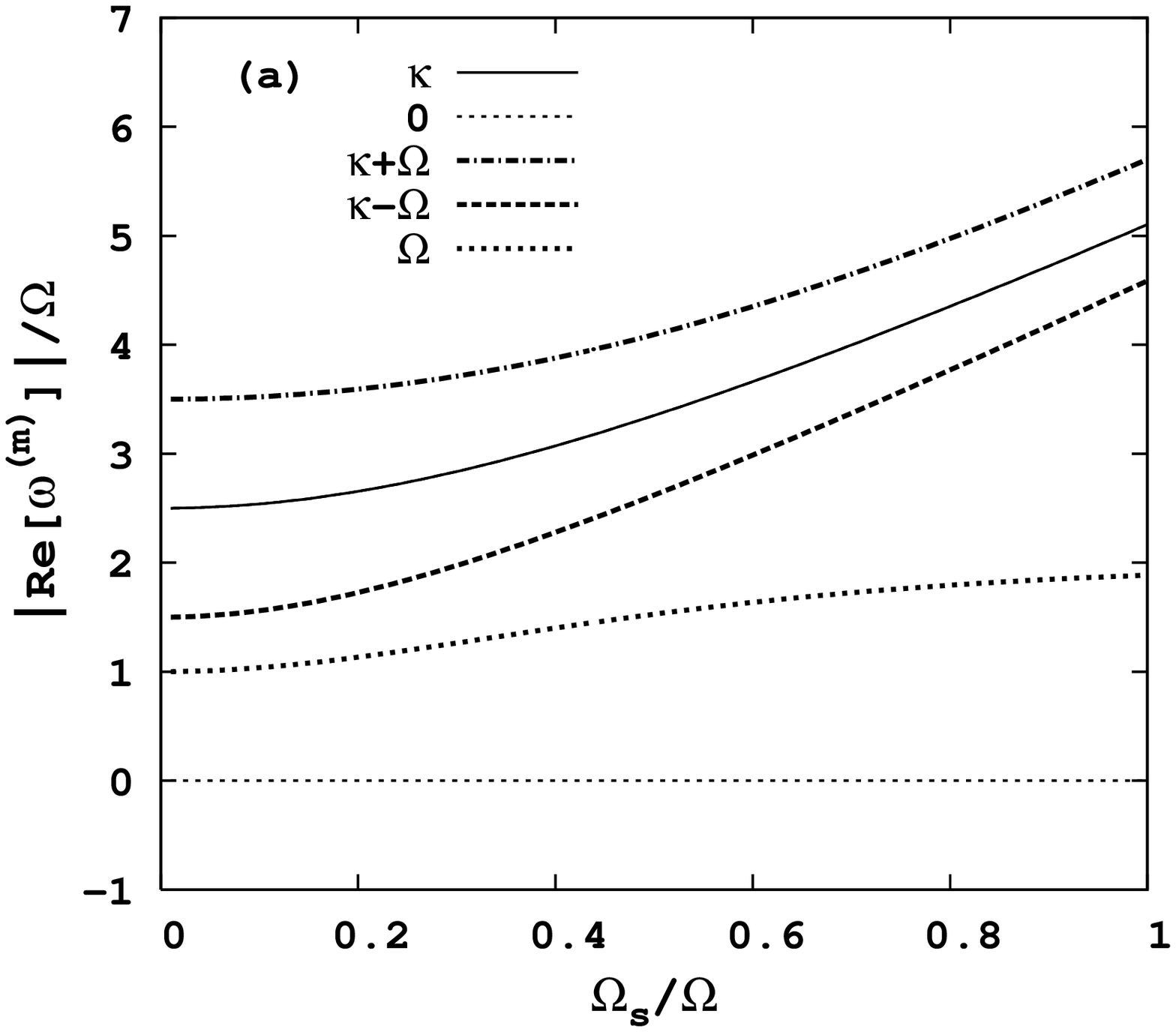}{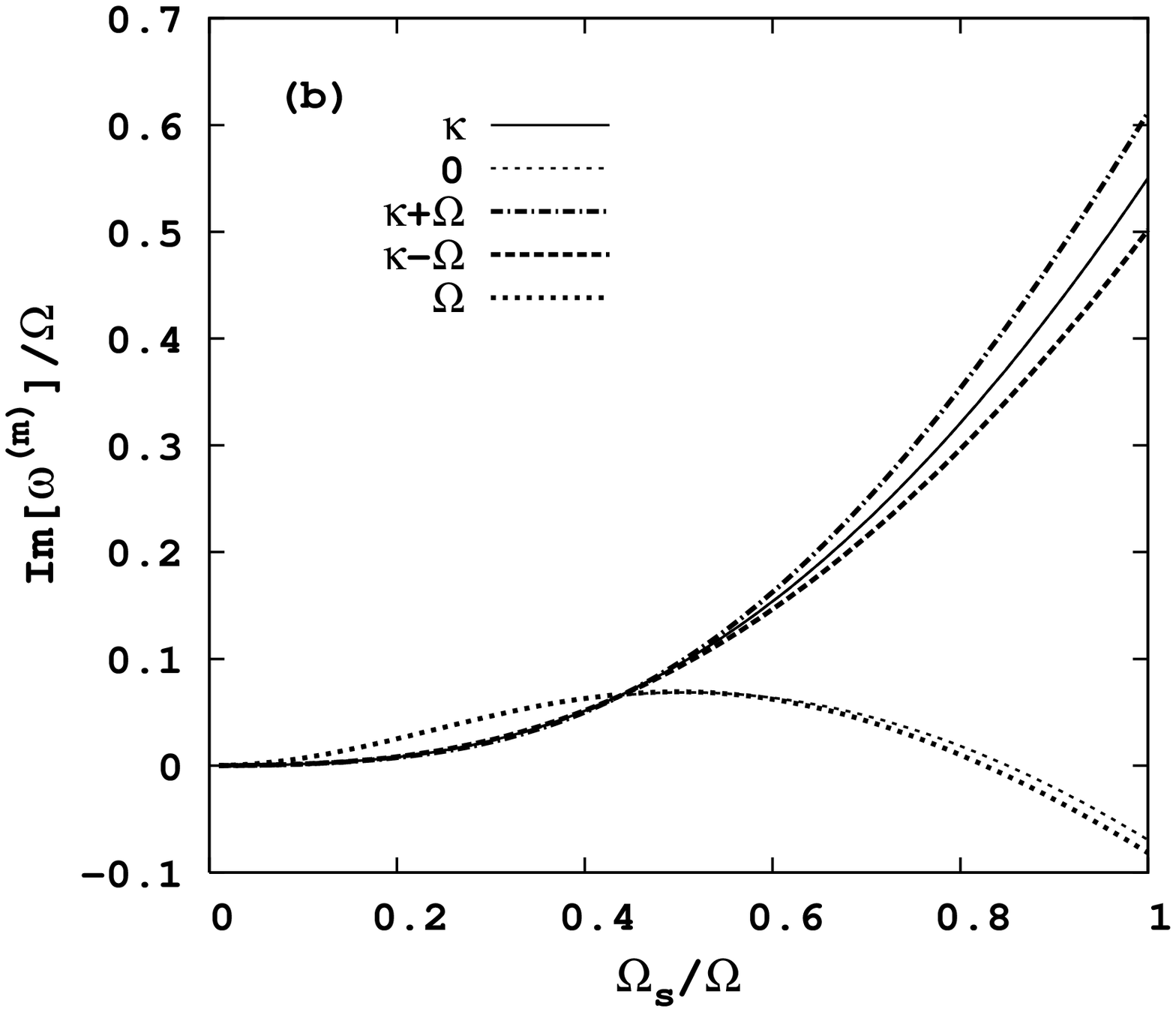} \caption{The real and
imaginary parts of the solutions to the dispersion relation for
axisymmetric ($m=0$) and non-axisymmetric ($m=1$) perturbations
excited at the innermost disk radius where $\Omega =\Omega _{*}$.
The model parameters are $\protect\kappa /\Omega =2.5$,
$\protect\beta =-5$, $f=1$, and $\langle F_{\protect\phi
}^{\mathrm{mag}}\rangle _{0}/v_{r0}\Omega =0$. The frequencies and
growth rates are obtained for $\protect\alpha =0.1$. The
growing modes are associated with the $\protect\kappa +\Omega $, $\protect%
\kappa $, and $\protect\kappa -\Omega $ branches, with $\protect\kappa %
+\Omega $ having the highest growth rate. The zero-frequency and
$\Omega $ modes grow for lower values of $\Omega _{s}/\Omega $, but
decay in the
regime of $\Omega _{s}/\Omega \lesssim 1$, where the $\protect\kappa $ and $%
\protect\kappa \pm \Omega $ modes have the fastest growth rates.
\label{fig3}}
\end{figure}


\begin{thebibliography}{99}
\bibitem[]{} Abramowicz, M. A., Karas, V., Klu{\'z}niak, W., Lee, W. H., \& Rebusco, P.
2003, \pasj, 55, 467

\bibitem[]{} Alpar, M. A., Hasinger, G., Shaham, J., \& Yancopoulos, S. 1992, \aap,
257, 627

\bibitem[]{} Alpar, M. A., \& Psaltis, D. 2005, preprint (astro-ph/0511412)

\bibitem[]{} Alpar, M. A., \& Y\i lmaz, A. 1997, NewA, 2, 225

\bibitem[]{} Erkut, M. H., \& Alpar, M. A. 2004, \apj, 617, 461

\bibitem[]{} Kato, S. 2001, \pasj, 53, 1

\bibitem[]{} M\'{e}ndez, M., \& Belloni, T. 2007, \mnras, 381, 790

\bibitem[]{} M\'{e}ndez, M., \& van der Klis, M. 1999, \apj, 517, L51

\bibitem[]{} Miller, M. C., Lamb, F. K., \& Psaltis, D. 1998, \apj, 508, 791

\bibitem[]{} Psaltis, D., Belloni, T., \& van der Klis, M. 1999, \apj, 520, 262

\bibitem[]{} Psaltis, D., \& Norman, C. 2000, preprint (astro-ph/0001391)

\bibitem[]{} Shakura, N. I., \& Sunyaev, R. A. 1973, \aap, 24, 337

\bibitem[]{} Stehle, R., \& Spruit, H. C. 1999, \mnras, 304, 674

\bibitem[]{} Stella, L., Vietri, M., \& Morsink, S. M. 1999, \apj, 524, L63

\bibitem[]{} van der Klis, M. 2000, \araa, 38, 717

\bibitem[]{} Wagoner, R. W. 1999, \physrep, 311, 259

\bibitem[]{} Wijnands, R., van der Klis, M., Homan, J., Chakrabarty, D.,
Markwardt, C. B., \& Morgan, E. H. 2003, \nat, 424, 44
\end{thebibliography}
\end{document}